\documentclass[10pt,conference]{IEEEtran}
\IEEEoverridecommandlockouts

\usepackage{cite}
\usepackage{amsmath,amssymb,amsfonts}
\usepackage{algorithmic}
\usepackage{graphicx}
\usepackage{textcomp}
\usepackage{xcolor}
\usepackage{booktabs}
\usepackage{multirow}
\usepackage{url}
\usepackage{float}

\def\BibTeX{{\rm B\kern-.05em{\sc i\kern-.025em b}\kern-.08em
    T\kern-.1667em\lower.7ex\hbox{E}\kern-.125emX}}

\begin{document}

\title{Ember: An Extensible Benchmark Suite for Quantum Annealing Embedding Algorithms
\thanks{This work was supported by the National Science Foundation under Grant Number 2442853.}}

\author{
\IEEEauthorblockN{Zachary Macaskill-Smith, Unmol Sharma,\\Melissa Warner, K\'alm\'an Varga}
\IEEEauthorblockA{
\textit{Department of Physics and Astronomy} \\
\textit{Vanderbilt University}\\
Nashville, TN, USA \\
\{zachary.n.macaskill-smith,unmol.sharma,\\melissa.a.warner,kalman.varga\}@vanderbilt.edu}
\and
\IEEEauthorblockN{David A.\ B.\ Hyde,~\IEEEmembership{Senior Member,~IEEE}}
\IEEEauthorblockA{
\textit{Department of Computer Science} \\
\textit{Vanderbilt University}\\
Nashville, TN, USA \\
david.hyde.1@vanderbilt.edu}
}

\maketitle

\begin{abstract}
Minor embedding is a required compilation step for quantum annealing, mapping logical problem graphs onto sparse hardware topologies.
Despite its central role in determining solution quality, no standardized benchmark exists for comparing embedding algorithms: prior studies use
incompatible graph libraries, inconsistent metrics, and non-reproducible
experimental setups, making cross-algorithm comparisons unreliable.
We present Ember (Embedding Minor Benchmark for Evaluative Reproducibility), 
an open-source benchmarking framework addressing this gap. Ember 
provides a standardized algorithm interface with seeded, 
reproducible execution infrastructure; a diverse graph library of 
24,016 instances spanning structured, random, and physics-motivated 
problem types not previously used in embedding benchmarks; and a 
unified analysis pipeline supporting all three current D-Wave 
hardware topologies (Chimera, Pegasus, Zephyr).
We evaluate five algorithms across the full library on Chimera and 
find that no algorithm dominates universally: rankings vary 
systematically with graph structure, and the best algorithm depends 
on the family being embedded. We also examine the effects of hardware topology (including Pegasus and Zephyr), qubit error rates, and evaluate a reinforcement-learning approach (\textsc{CHARME}) within a narrower test set. 
Ember is available at \url{https://github.com/zachmacsmith/ember} and is installable via \texttt{pip install ember-qc}.
\end{abstract}


\section{Introduction}

Quantum annealing (QA) solves combinatorial optimisation problems by
encoding them as Ising Hamiltonians and exploiting quantum fluctuations to
find low-energy configurations~\cite{kadowaki1998}. Current D-Wave
processors implement QA on sparse hardware graphs (Chimera, Pegasus, and
Zephyr) whose limited qubit connectivity requires that every problem be
compiled via \emph{minor embedding}: a mapping from logical problem
variables onto physical qubit chains such that the problem's interaction
graph is realised as a minor of the hardware graph~\cite{choi2008}.

Embedding quality directly constrains annealing performance. Longer qubit
chains increase the likelihood of chain breaks during annealing, degrade
solution fidelity, and consume additional qubits that could otherwise
encode larger problems~\cite{gomez2025}. Finding a high-quality embedding
is therefore critical to practical quantum annealing, and embedding is
often the dominant cost in the problem compilation pipeline.

Despite this centrality, no standardized benchmark exists for comparing
embedding algorithms. The six primary algorithms in use today (MinorMiner~\cite{cai2014}, OCT-based embedding~\cite{date2019},
ATOM~\cite{ngo2023atom}, PSSA~\cite{sugie2018,sugie2021},
Clique embedding~\cite{boothby2016}, and CHARME~\cite{ngo2024charme}) have each been evaluated in isolation, on incompatible experimental setups:
\begin{itemize}
  \item ATOM evaluates on Barab\'{a}si-Albert and $d$-regular graphs on
    Chimera only, with no random seed reported~\cite{ngo2023atom}.
  \item CHARME evaluates on Barab\'{a}si-Albert graphs on Chimera only,
    with 600 synthetic test instances~\cite{ngo2024charme}.
  \item The most extensive existing comparison~\cite{zbinden2020} covers
    Erd\H{o}s-R\'{e}nyi, Barab\'{a}si-Albert, and $d$-regular graphs on
    Chimera and Pegasus, but predates both ATOM and CHARME.
  \item No study covers Zephyr as a target topology in an
    algorithm comparison.
\item The most recent embedding comparison,
    Pelofske et al.~\cite{pelofske2024scaling}, benchmarks
    \textsc{MinorMiner} and \textsc{Clique} across all three
    topologies with analytic scaling models, but does not
    evaluate \textsc{OCT}, \textsc{ATOM}, \textsc{PSSA}, or
    \textsc{CHARME}.
\end{itemize}
These incompatibilities mean that, for instance, CHARME's claim to outperform ATOM cannot
be verified against the same conditions under which ATOM was evaluated.
More generally, a
researcher seeking to choose an embedding algorithm for a specific problem
class has no reliable basis for comparison.

This paper presents Ember, addressing these gaps with four primary
contributions:

\begin{enumerate}
  \item \textbf{Reproducible benchmarking infrastructure.} A 
    standardized algorithm interface with seeded execution, 
    resolved configuration saved alongside results, and 
    version-tracked outputs. Every benchmark run is fully 
    reproducible from a single YAML experiment file.

  \item \textbf{A diverse graph library.} 24,016 instances spanning 
    35 graph categories across five structural families, including 
    physics-motivated lattices (triangular, kagome, frustrated 
    square) and benchmarking instances (planted-solution, 
    weak-strong cluster, spin-glass) not previously used in 
    embedding benchmarks. All three current D-Wave hardware 
    topologies are supported as targets.

\item \textbf{Cross-algorithm comparison revealing structural 
    rank reversals.} We evaluate five algorithms across 17,248 
    Chimera-embeddable instances and find that no algorithm 
    dominates universally: rankings vary systematically with 
    graph family, and the best algorithm depends on the structural 
    properties of the input. Aggregate rankings ($W = 0.647$) mask 
    these reversals.

\item \textbf{Hardware topology and fault-tolerance 
characterisation.} We characterise embedding-capacity 
differences across all three D-Wave topologies, finding that 
Zephyr extends Pegasus's edge capacity by 12\% while 
Chimera's lower connectivity bounds practical problem size 
at roughly $3\times$ less. Under simulated qubit faults across 
a range from current hardware levels to severe degradation 
($f \in [0, 0.25]$), \textsc{MinorMiner} degrades gracefully 
while \textsc{PSSA} and \textsc{Clique} exhibit sharp cliffs, 
losing roughly twice \textsc{MinorMiner}'s success rate at 
the high end of the tested range.
\end{enumerate}

Ember is open source, pip-installable (\texttt{pip install ember-qc}),
and provides a published algorithm contract enabling future algorithms to
report results on the same benchmark.

\section{Background}

\subsection{Minor Embedding}

Let $G = (V_G, E_G)$ be a logical problem graph and $H = (V_H, E_H)$ be
a hardware graph. A \emph{minor embedding} of $G$ into $H$ is a mapping
$\phi: V_G \rightarrow 2^{V_H}$ such that: (i) each image $\phi(v)$
(called a \emph{chain}) induces a connected subgraph of $H$; (ii) chains
are vertex-disjoint; and (iii) for every edge $(u,v) \in E_G$, there
exists at least one edge in $H$ between $\phi(u)$ and $\phi(v)$.

Finding a minor embedding is NP-hard in general~\cite{cai2014}. Because
each logical variable is represented by a chain of physical qubits, chain
length directly affects annealing fidelity: longer chains require stronger
ferromagnetic coupling to maintain chain integrity, leaving less dynamic
range for problem-encoding couplers. Chain length is therefore the primary
quality metric for embeddings.

\subsection{Hardware Topologies}

We benchmark across three hardware topologies corresponding to current and
recent D-Wave processors:

\begin{itemize}
  \item \textbf{Chimera $C_{16}$}: 2,048 nodes, 6,016 edges. The legacy
    topology present in D-Wave 2000Q devices. Bipartite unit cells of
    $K_{4,4}$.
  \item \textbf{Pegasus $P_{16}$}: 5,640 nodes, 40,484 edges. Present in
    D-Wave Advantage processors. Higher connectivity than Chimera with
    degree-15 nodes \cite{boothby2020pegasus, dwaveDocsPegasus}.
  \item \textbf{Zephyr $Z_{12}$}: 4,800 nodes, 45,864 edges. Present in
    D-Wave Advantage2 processors. Highest edge count of the three,
    enabling shorter chains for dense source graphs ~\cite{dwaveDocsZephyr}.
\end{itemize}

\subsection{Embedding Quality Metrics}

We report the following metrics per trial:

\begin{itemize}
  \item \textbf{Success rate}: fraction of trials in which a valid
    embedding was found.
  \item \textbf{Maximum chain length}: length of the longest chain in
    the embedding.
  \item \textbf{Mean chain length}: average chain length across all
    logical variables.
  \item \textbf{Wall time}: elapsed time measured externally around the
    embedding call.
\end{itemize}

\section{Related Work}

\subsection{Embedding Algorithms}
\label{sec:algorithms}

\textbf{MinorMiner}~\cite{cai2014, minorminerSoftware} is the standard 
heuristic provided in
D-Wave's Ocean software stack~\cite{oceanSoftware}. It iteratively constructs chains by finding
shortest paths in the hardware graph, refining an initially infeasible
solution until a valid embedding is found or the timeout is reached. It is
the de facto baseline and is included in all subsequent comparisons.

\textbf{Clique embedding}~\cite{boothby2016} provides exact, polynomial-time
embeddings for complete graphs by exploiting the structure of Chimera,
Pegasus and Zephyr topologies. It is optimal for dense logical graphs, but
limited to complete-graph inputs; its clique embedding is commonly used as
an all-to-all embedding for arbitrary problems, at the cost of increased
chain length.

\textbf{PSSA} (Probabilistic Swap-Shift
Annealing)~\cite{sugie2018,sugie2021} applies simulated annealing
to the embedding problem. The improved version was shown to
consistently exceed the complete-graph embedding threshold by
factors of 3.2 and 2.8 on random cubic and Barab\'{a}si-Albert
graphs respectively, on hardware graphs up to 102,400 nodes.

\textbf{ATOM}~\cite{ngo2023atom} uses an adaptive topology approach
to find embeddings with shorter runtime than OCT-based methods while
matching their qubit usage. Evaluation covers Barab\'{a}si-Albert and
$d$-regular graphs on Chimera, comparing favorably against MinorMiner
and OCT-based embedding on runtime.

\textbf{OCT-based embedding}~\cite{date2019} achieves the highest known
embedding quality among current algorithms by exploiting odd-cycle
transversal (OCT) decompositions. It is computationally expensive, often
by an order of magnitude over MinorMiner, and was originally developed
for Chimera topologies.

\textbf{CHARME}~\cite{ngo2024charme} introduces a reinforcement learning
approach using graph neural networks to learn an ordering policy for
logical variable placement. It outperforms MinorMiner and ATOM on qubit
usage for sparse graphs, and in some cases, surpasses OCT-based embedding.
Evaluation is conducted on Barab\'{a}si-Albert graphs on Chimera.

Additional approaches include integer-programming
formulations~\cite{bernal2020}, layout-aware
embedding~\cite{pinilla2019}, Spring-Based and Clique-Based
MinorMiner variants~\cite{zbinden2020}, alternative
chain-length-minimising heuristics~\cite{yang2019},
deterministic bipartite-graph
embeddings~\cite{sinno2025}, and a recent reinforcement
learning approach on Zephyr topologies~\cite{nembrini2025}.
Caching strategies for accelerating embedding were contemplated
by Brahm et al.~\cite{brahm2021caching}, and further deep
learning approaches by Hyde~\cite{hyde2025deep}. Embedding
algorithms remain an open research area, and it is necessary
to have an open and extensible benchmark that new algorithms
can easily evaluate against.

\subsection{The Reproducibility and Comparability Gap}
The embedding algorithms enumerated above cannot be directly
compared. Each has been evaluated on a different graph
distribution, hardware topology, and experimental protocol:
ATOM evaluates on Barab\'{a}si-Albert and $d$-regular graphs
on Chimera only; CHARME on Barab\'{a}si-Albert graphs on
Chimera only; neither reports random seeds. PSSA evaluates
primarily on Hitachi's CMO architecture using King's-graph
hardware up to 102,400 nodes~\cite{sugie2021}, with only a
preliminary Chimera adaptation~\cite{sugie2018}. The
additional approaches listed above follow the same pattern:
each evaluates on its own combination of graph families
(random, lattice-structured, or single graph classes) and
topology subset (Chimera, Pegasus, or both), with no overlap
enabling direct comparison.

Partial comparisons exist but each covers a narrow slice.
The most systematic prior algorithm comparison is that of
Zbinden et al.~\cite{zbinden2020}, who benchmark four
algorithms on Erd\H{o}s-R\'{e}nyi, Barab\'{a}si-Albert, and
$d$-regular graphs across Chimera and Pegasus, but this
predates ATOM and CHARME by three years and does not cover
Zephyr. G\'{o}mez-Tejedor and Osaba~\cite{gomez2025} evaluate
MinorMiner and Clique embedding on Pegasus, finding
significant room for improvement, but do not evaluate other
algorithms or provide a reproducible framework. Pelofske et
al.~\cite{pelofske2024scaling} extend to all three topologies
with analytic scaling models, but again benchmark only
MinorMiner and Clique. No existing work places all major
competitive algorithms on a common graph library with an 
open-source, reproducible evaluation protocol.

\subsection{Graph Library Diversity}

All existing embedding benchmarks use only three random graph families:
Erd\H{o}s-R\'{e}nyi $G(n,p)$, Barab\'{a}si-Albert, and $d$-regular
graphs. No prior embedding study has systematically evaluated algorithm
performance on structured graph types (grids, cycles, complete graphs,
bipartite graphs) or on physics-motivated lattice graphs, despite the fact
that these arise directly in quantum simulation workloads.
For instance, King et al.~\cite{king2023} demonstrated quantum simulation of frustrated magnets on Kagome-structured Hamiltonians using D-Wave hardware, yet no benchmark has tested whether embedding algorithms handle Kagome source graphs differently from random graphs of equivalent size and density.

\subsection{Quantum Annealing Benchmarking}

Broader quantum annealing benchmarking work (e.g.,  Pelofske~\cite{pelofske2025}, Vert et al.~\cite{vert2024}) focuses on annealing solution quality rather than embedding algorithm comparison. This work is complementary: Ember isolates the embedding step and does not depend on QPU access, enabling classical evaluation of the compilation pipeline.

\section{The Ember Framework}

\subsection{Architecture Overview}

Ember consists of two packages: \texttt{ember-qc}, the benchmark runner,
and \texttt{ember-qc-analysis}, a standalone analysis package with no
quantum dependencies. The runner handles experiment execution, result
collection, and checkpointing; the analysis package handles plotting,
statistics, and tabular summaries. The two-package design allows analysis
to be performed on any machine, including those without the D-Wave software
stack.

The benchmark runner pipeline is: (i) parse experiment configuration from
YAML; (ii) build task list from graph selection and algorithm list; (iii)
dispatch tasks to parallel workers; (iv) collect and validate results;
(v) write structured output, including the fully resolved configuration for
reproducibility.

\subsection{Algorithm Interface}

Every embedding algorithm in Ember is a Python class subclassing
\texttt{EmbeddingAlgorithm} and decorated with
\texttt{@register\_algorithm}. The interface requires:
\begin{enumerate}
  \item An \texttt{embed(source\_graph, target\_graph, **kwargs)} method
    returning \texttt{\{`embedding': \{node: [chain]\}\}} on success or
    \texttt{\{`embedding': \{\}\}} on failure. \texttt{None} is never
    returned.
  \item A \texttt{version} property returning a version string.
  \item Respect for the \texttt{timeout} passed via \texttt{kwargs}.
  \item No modification of input graphs, which are shared across parallel
    workers.
\end{enumerate}

The runner measures wall time externally around the
\texttt{embed()} call, infers success from embedding non-emptiness, and
validates every returned embedding structurally.
Algorithms may optionally report a failure status (\texttt{TIMEOUT}, \texttt{FAILURE}, \texttt{OOM}) and algorithmic counters for cross-algorithm comparison.
This interface is published as a stable contract, enabling future algorithm implementations to report results on Ember directly.

\subsection{Reproducibility Infrastructure}

Reproducibility is a primary design goal. Ember provides:

\begin{itemize}
  \item \textbf{Seeded execution.} A master seed is set per run; per-trial
    seeds are derived deterministically. The runner reseeds Python's
    \texttt{random} and \texttt{numpy.random} before each trial, and
    passes the seed to algorithms via \texttt{kwargs}.
  \item \textbf{Resolved configuration.} The fully resolved experiment
    configuration, including all flag overrides, Ember version, and
    timestamp, is saved as \texttt{experiment\_resolved.yaml} in the
    output directory alongside results.
  \item \textbf{Checkpoint and resume.} Interrupted benchmark runs can be
    resumed from the point of interruption, preserving completed trials.
  \item \textbf{Faulty qubit simulation.} Hardware faults can be simulated
    by removing nodes and edges from target graphs at a specified fault
    rate, enabling evaluation under realistic hardware conditions.
\end{itemize}

\subsection{Graph Library}
The Ember library comprises 24{,}016 distinct problem graphs drawn from 35 categories across 5 families, spanning 2 to 65{,}536 nodes. It extends prior embedding
benchmarks in two main ways: it includes structured families that arise in
specific quadratic unconstrained binary optimisation (QUBO) problem classes (complete graphs in portfolio optimisation
and max-cut; bipartite graphs in matching and assignment; grid graphs in
image segmentation) but are absent from prior benchmarks, and it includes physics-motivated lattice types whose structural properties (regularity, clustering coefficient, treewidth) produce qualitatively different embedding difficulty profiles across algorithms. Several niche families, including topological extremals chosen for treewidth and degree-sequence properties, application-derived benchmarks with known ground states (planted-solution and weak-strong cluster instances), and hardware-native subgraphs that anchor a lower bound on embedding cost, provide complementary reference points for differentiating algorithms across structurally distinct regimes. 

\paragraph{Random (17{,}567 graphs).}
Watts--Strogatz~\cite{watts1998} (9{,}760), Erd\H{o}s--R\'{e}nyi~\cite{erdos1959} (2{,}395),
Barab\'{a}si--Albert~\cite{barabasi1999} (2{,}263), $d$-regular (2{,}273), stochastic block
model~\cite{holland1983} (612), random planar (213), and LFR benchmark graphs~\cite{lancichinetti2008} (51) provide
broad structural and density coverage.

\paragraph{Structured (1{,}557 graphs).}
Algebraically defined graphs with predictable regularity. Generalised
Petersen (496) and circulant graphs (351) are vertex-transitive, probing
exploitation of regularity. Tur\'{a}n graphs $T(n,r)$ (397) and complete
bipartite graphs (147) cover extremal density regimes. Complete graphs
$K_n$ (56) are maximally dense worst cases for qubit overhead. Johnson
(63), Kneser (36), and hypercube graphs (11) have well-studied chromatic
and dimensional properties.

\paragraph{Physics / Lattice (880 graphs).}
Standard 2-D lattices (triangular (194), honeycomb (141), square grid
(101)) and frustrated variants (kagome (126), king (72), frustrated
square (72), Shastry--Sutherland (72)) model spin systems of direct
experimental relevance; triangular and kagome lattices arise
specifically in frustrated magnet and spin-ice simulation. 3-D crystal
structures (cubic 76; BCC 26) extend coverage to less regular spatial
geometries.

\paragraph{Topological (286 graphs).}
Paths, cycles, stars, and wheels (64, 64, 63, 63) are extremal in
treewidth, diameter, or degree sequence. Balanced trees (21) and binary
trees (11) cover hierarchical structure.

\paragraph{Benchmarking / Application (3{,}726 graphs).}
Planted-solution graphs (2{,}616) guarantee feasibility and supply a
ground-truth quality reference. Weak-strong cluster graphs~\cite{denchev2016} (456) are an
established QUBO benchmark with a known ground state easily obscured by
local optima. Spin-glass instances (598) provide disordered reference
problems. Hardware-native subgraphs (42) are induced subgraphs of the
target topologies, providing a lower-bound reference on embedding
difficulty. Named special graphs (12) include well-known instances such
as the Petersen and Heawood graphs.

\paragraph{CHARME training/evaluation suite (45 graphs)}
A separate set of 45 Chimera-only graphs at $n=120$, used to evaluate
\textsc{CHARME} at its trained action-space size without padding. The
suite spans the same families: 39 random graphs (Erd\H{o}s--R\'{e}nyi
12, Watts--Strogatz 10, Barab\'{a}si--Albert 8, $d$-regular 3,
stochastic block model 3, random planar 3), 3 physics/lattice graphs
(grid), and 3 topological graphs (path, cycle, star). Including this
suite, the library totals 24{,}061 distinct graphs.

Instance counts per category reflect the dimensionality of each
generator's parameter space: families with more free parameters
(e.g., Watts--Strogatz with $n$, $k$, and $\beta$) require more
instances to cover their structural range than single-parameter
families (e.g., complete graphs $K_n$). To prevent high-instance
categories from dominating aggregate statistics, all cross-category
comparisons use macro-averaging over categories rather than pooling
instances. Of the 24{,}016 distinct graphs, 17{,}260 fit within Chimera~$C_{16}$; 23{,}652 within Pegasus~$P_{16}$; and 23{,}131 within Zephyr~$Z_{12}$, yielding 64{,}043 graph--topology pairs across the full benchmark.



\section{Experimental Setup}

\subsection{Algorithms Evaluated}
\label{sec:algorithms_evaluated}

We evaluate the following algorithms, all implemented using Ember's
algorithm contract:

\begin{itemize}
  \item \textbf{MinorMiner} (default, greedy, and systematic 
    variants), version 0.2.21, from D-Wave's \texttt{minorminer} 
    package~\cite{minorminerSoftware}.
    
  \item \textbf{Clique embedding} via \texttt{minorminer.busclique} \cite{minorminerSoftware}.
  
  \item \textbf{ATOM}, binary build adapted from the authors' 
    released code\footnote{\url{https://github.com/ngominhhoang/Quantum-annealing-minor-embedding}}. 
    Chimera only.
  
  \item \textbf{OCT-based} (fast-OCT and hybrid-OCT variants), 
    based on the authors' released 
    code\footnote{\url{https://github.com/TheoryInPractice/aqc-virtual-embedding}}. 
    Chimera only.
    \item \textbf{PSSA} (standard, weighted, fast, and thorough 
variants), custom implementation. The original PSSA was developed for Hitachi's CMO
architecture using King's-graph
adjacency~\cite{sugie2018,sugie2021}; a preliminary
Chimera adaptation in~\cite{sugie2018} matched but did
not exceed \textsc{MinorMiner}, and its authors noted
further adaptation as future work. To our knowledge, no publicly available implementation generalised PSSA to all
D-Wave hardware topologies prior to this work.

We reimplemented 
the simulated-annealing search and terminal-search 
post-processing from the original paper, with three 
substantive adaptations: the King's-graph guiding-path 
initialisation is replaced with a busclique-derived clique 
embedding on the D-Wave hardware graph, ensuring a feasible 
starting state whenever the source graph fits the hardware's 
maximum embeddable clique; chain-mutation primitives 
operate on tree-shaped chains (the form busclique produces) 
rather than the simple paths of the original, with 
connectivity preserved via induced-subgraph checks; and 
annealing schedule parameters are scaled per-topology to 
account for the qubit-degree differences between Chimera, 
Pegasus, and Zephyr. The four registered variants differ 
in shift-proposal weighting and annealing budget.

\item \textbf{CHARME}~\cite{ngo2024charme}, a reinforcement 
    learning approach using a GCN-based actor-critic policy to 
    learn an ordering over logical nodes for embedding. The 
    original authors did not release pretrained weights; we 
    reconstructed the training pipeline from the released 
    code\footnote{\url{https://github.com/ngominhhoang/charme-rl-minor-embedding}}. We trained CHARME on a diverse corpus of 150 graphs drawn 
from the same families as the Ember library, with disjoint 
test instances, in contrast to the original paper's exclusively 
Barab\'asi--Albert training. All training graphs were fixed at 
$n = 120$, as the architecture requires a fixed action dimension 
at initialisation; variable-size graphs produce incompatible tensor 
shapes during batched proximal policy optimization~\cite{schulman2017ppo} (PPO) updates. We extend the inference pipeline 
with zero-padding to evaluate smaller graphs at test time, though 
the policy itself was optimised only at $n = 120$. Training and inference used a 
Chimera($16,16,4$) hardware graph rather than the paper's 
$45 \times 45$ grid, matching deployed D-Wave 2000Q hardware. 
Results should be interpreted as indicative rather than definitive 
pending longer training runs.
    
\end{itemize}
All algorithm implementations used in this study, including
our \textsc{PSSA} reimplementation and \textsc{CHARME}
training pipeline, are publicly available alongside the
benchmark at \url{https://github.com/zachmacsmith/ember}.

\subsection{Graph Selection}

The main comparison evaluates against the full Ember library of 
24,016 graph instances using the \texttt{default} preset, run against 
all three hardware topologies: Chimera $C_{16}$, Pegasus $P_{16}$, 
and Zephyr $Z_{12}$. Two further analyses, variant selection (\textbackslash{}S\textbackslash{}ref\{sec:variant\_selection\}) and fault tolerance (\textbackslash{}S\textbackslash{}ref\{sec:fault\}), use a 273-graph \textbackslash{}texttt\{sensitivity\} subset that covers all graph categories across their difficulty ranges, selected for analyses where compute cost per graph is substantially higher than in the main comparison. 

\subsection{Experimental Parameters}
Each algorithm--graph--topology combination runs with a 30\,s 
per-trial timeout; master seed fixed at 42. \textsc{OCT-fast} uses 
$r = 100$ restarts (sensitivity validated in \S\ref{sec:variant_selection}).
All experiments were run on a server equipped with two AMD EPYC 75F3 
32-Core processors (128 total threads), 512\,GB RAM, and dual NVIDIA 
RTX A6000 GPUs (48\,GB VRAM each), running Debian GNU/Linux 12 
(Bookworm) with CUDA 12.2. Statistical significance is assessed 
using the Wilcoxon signed-rank test with Holm--Bonferroni correction 
for multiple comparisons; overall ranking differences are evaluated 
via the Friedman test.

\subsection{Algorithm Variant Selection}
\label{sec:variant_selection}

\textsc{MinorMiner}, \textsc{PSSA}, and \textsc{OCT} each expose 
multiple implementation variants.
We adopted a two-stage experimental design: first, all variants were evaluated on a sensitivity-graph 
subset, and then the best variant of each method was used in subsequent tests. 
The default variant in a given library was used when there was no clear winner.

No \textsc{MinorMiner} variant improved on the default by more
than 1.3~pp success; \textsc{PSSA}'s four variants collapsed to
within 0.3~pp success and 0.05 qubits ACL. In both cases we
advanced the default.

\textsc{OCT} was the opposite case. Its six variants 
(\texttt{triad}, \texttt{triad-reduce}, \texttt{hybrid-oct}, 
\texttt{hybrid-oct-reduce}, \texttt{fast-oct}, 
\texttt{fast-oct-reduce}) are structurally distinct decomposition 
and chain-construction procedures rather than parameter settings 
of a shared core, and performance differences reflected this: 
success rate ranged from 3.6\% (\texttt{triad-reduce}) to 48.6\% 
(\texttt{fast-oct-reduce}), and mean ACL from 2.00 to 7.74. 
\texttt{fast-oct-reduce} outperformed the nearest alternative on 
both metrics (48.6\% / 3.70 ACL vs.\ 47.8\% / 5.47 for 
\texttt{fast-oct}) at negligible runtime cost.
We advanced \texttt{fast-oct-reduce} as 
\textsc{OCT-fast} throughout \S\ref{sec:results}. Restart-count sensitivity was 
separately validated on the full library ($r \in \{100, 1{,}000, 
10{,}000,$ adaptive$\}$; success rate varies by 0.2 pp, ACL by 0.02), 
confirming that restart count is negligible relative to structural 
configuration.

The order-of-magnitude spread across \textsc{OCT}'s structural 
configurations suggests unexplored design space within the 
\textsc{OCT} family: the selected variant is not obviously 
optimal, and further structural refinement could plausibly 
improve \textsc{OCT}'s position in the cross-algorithm rankings.

\section{Results}
\label{sec:results}

\subsection{Overall Algorithm Comparison}
\label{sec:overall}

Table~\ref{tab:overall} presents the mean success rate, mean average chain length (ACL), and median ACL across all graph types.
\begin{table}[htbp]
\centering
\caption{Overall algorithm performance on the Chimera~16$\times$16$\times$4 benchmark
(17\,248 problem graphs), macro-averaged across graph categories.
Success rate and mean ACL shown as mean~$\pm$~95\%~CI.
Median wall time is the macro-average of per-category medians
(successful embeddings only); its 95\% CI is from a within-category
bootstrap (2000 resamples).}
\label{tab:overall}
\begin{tabular}{lccc}
\toprule
\textbf{Algorithm} & \textbf{Success Rate} & \textbf{Mean ACL} & \textbf{Median Time (s)} \\
\midrule
MinorMiner & \textbf{61.3 $\pm$ 2.2\%} & \textbf{4.23 $\pm$ 0.20} & 0.771 $\pm$ 0.176 \\
OCT-fast   & 42.1 $\pm$ 2.2\%          & 4.37 $\pm$ 0.17          & 0.091 $\pm$ 0.023 \\
PSSA       & 39.8 $\pm$ 2.2\%          & 5.37 $\pm$ 0.22          & 0.258 $\pm$ 0.452 \\
ATOM       & 36.8 $\pm$ 2.3\%          & 5.10 $\pm$ 0.25          & 0.028 $\pm$ 0.004 \\
Clique     & 36.7 $\pm$ 2.2\%          & 9.02 $\pm$ 0.38          & \textbf{0.004 $\pm$ 0.000} \\
\bottomrule
\end{tabular}
\end{table}

\begin{table*}[t]
\centering
\small
\caption{Categories where the global ranking (\textsc{MM} $>$
\textsc{OCT} $>$ \textsc{PSSA} $>$ \textsc{ATOM} $>$ \textsc{Clique})
does not hold at the top-two positions. Bold = top rank.
All Friedman tests $p < 0.001$.}
\label{tab:category_exceptions}
\begin{tabular}{llrr rrrrr}
\toprule
\textbf{Family} & \textbf{Category} & \textbf{$N$} & \textbf{$W$} &
\textbf{MM} & \textbf{OCT} & \textbf{PSSA} & \textbf{ATOM} & \textbf{Clique} \\
\midrule
Random          & Erd\H{o}s--R\'enyi   & 1,245 & 0.433 & \textbf{1.63} & 2.70 & 2.64 & 4.33 & 3.69 \\
\addlinespace
Structured      & Bipartite            &    79 & 0.692 & 2.01 & \textbf{1.62} & 2.63 & 4.08 & 4.66 \\
                & Complete ($K_n$)     &    30 & 0.563 & 3.32 & 2.00 & \textbf{1.82} & 4.77 & 3.10 \\
                & Hypercube            &     6 & 0.810 & 1.92 & \textbf{1.25} & 3.50 & 3.50 & 4.83 \\
                & Kneser               &    12 & 0.546 & 2.33 & \textbf{1.42} & 3.04 & 4.29 & 3.92 \\
                & Tur\'an              &   249 & 0.635 & 2.82 & \textbf{1.38} & 2.37 & 4.63 & 3.80 \\
\addlinespace
Physics         & Kagome               &    31 & 0.618 & \textbf{1.00} & 3.23 & 3.81 & 2.77 & 4.19 \\
\addlinespace
Topological     & Star                 &    42 & 0.512 & \textbf{1.95} & 2.67 & 1.99 & 4.10 & 4.30 \\
\addlinespace
Benchmarking    & Planted solution     & 1,357 & 0.506 & \textbf{1.00} & 3.49 & 3.30 & 3.61 & 3.60 \\
                & Spin glass           &   282 & 0.345 & 2.35 & \textbf{2.24} & 2.50 & 4.39 & 3.52 \\
                & Weak/strong cluster  &   101 & 0.485 & \textbf{1.13} & 3.40 & 2.99 & 3.52 & 3.96 \\
\bottomrule
\end{tabular}
\end{table*}

Restricting to the 5{,}215 problems on which all five algorithms succeeded,
the Friedman test confirms significant differences in chain-length ranking
($\chi^2 = 13{,}503.1$, $p < 0.001$, $W = 0.647$), with mean ranks placing
\textsc{MinorMiner} first (1.54), followed by \textsc{OCT-fast} (1.89),
\textsc{PSSA} (3.21), \textsc{ATOM} (3.83), and \textsc{Clique} (4.52);
all pairwise differences are significant ($p < 0.001$ across all
$\binom{5}{2} = 10$ comparisons). Extending to all 10\{,\}381 problems on which at least one algorithm succeeded, with failed embeddings ranked last and tied where multiple algorithms fail on the same problem, the ranking order is unchanged but the gaps widen ($\chi^2 = 19{,}561.9$, $p < 0.001$, $W =
0.471$): \textsc{OCT-fast} falls further behind \textsc{MinorMiner} ($+0.63$)
once its 19~pp lower success rate is penalised, while \textsc{Clique} improves
($-0.44$) because its failures tend to co-occur with failures by other
algorithms, distributing the penalty equally. The drop in $W$ from $0.647$
to $0.471$ reflects greater cross-algorithm disagreement over the full problem
set, where varying failure patterns introduce additional rank inconsistency.

The moderate Kendall's $W = 0.647$ indicates that while a consistent
overall ranking exists, substantial variation in relative algorithm
performance across graph categories prevents any single algorithm from
dominating universally. The runtime data in Table~\ref{tab:overall}
sharpens this picture: when \textsc{OCT-fast} succeeds, it achieves
chain lengths within 3\% of \textsc{MinorMiner} at roughly one-eighth the
runtime (0.091\,s vs.\ 0.771\,s), making success rate the primary
differentiator between the two. \textsc{ATOM} was designed as a
faster alternative to \textsc{MinorMiner} with comparable quality,
but across this benchmark its $\sim$28$\times$ speed advantage
(0.028\,s) comes with both lower success rate and worse chain
quality, suggesting its design assumptions favour a narrower graph
distribution than tested here. \textsc{Clique} is fastest by a
large margin (0.004\,s) but produces chains more than twice as long
on average, confirming it is best-suited for when speed
constraints strongly dominate.

\subsection{Performance by Graph Family}
\label{sec:by_family}
To test whether algorithm rankings vary significantly across graph
structures, we apply the penalised Friedman test separately within
each of the 35 graph categories, with failures ranked last. The ART
ANOVA interaction term~\cite{wobbrock2011} ($F(136, 25{,}900) = 8.18$, $p < 10^{-148}$)
confirms that ranking differences between algorithms depend on which
graph type is being embedded; family-level aggregation masks this
heterogeneity (Benchmarking family mean $W = 0.520$) by pooling
categories whose rankings actively contradict each other.

At the category level ($n = 35$ types, mean $W = 0.611$), ranking
differences between algorithms depend systematically on graph
structure. \textsc{MinorMiner} dominates every Physics-lattice
category and every Topological-primitive
category (binary tree, cycle, path, star, tree, wheel; all top rank).
\textsc{OCT-fast} leads on algebraically regular structured graphs (bipartite, Tur'an, Kneser, hypercubes) and on spin-glass instances, suggesting an advantage on
degree-homogeneous or highly symmetric graphs. \textsc{PSSA} ranks
first on only one category, Complete graphs---its sole top-ranked
category, characterised by uniform maximal edge density.

Furthermore, these rank reversals correspond to practically significant
chain-length differences. On the intersection of graphs solved by both
algorithms, \textsc{OCT-fast} produces chains 7.8\% shorter than
\textsc{MinorMiner} on Tur\'an graphs, 15.8\% shorter on Kneser, and
5.4\% shorter on bipartite graphs and on spin-glass instances. On
Complete graphs \textsc{PSSA} ties \textsc{OCT-fast} within ${\sim}2\%$
(slightly longer) but leads \textsc{MinorMiner} by 13.9\%.

This demonstrates that benchmarking exclusively on random graphs,
as in prior work, gives an incomplete picture of algorithm behaviour
on the graph types arising in practical quantum annealing workloads.

\begin{figure}[hbtp]
\centering
\includegraphics[width=\columnwidth]{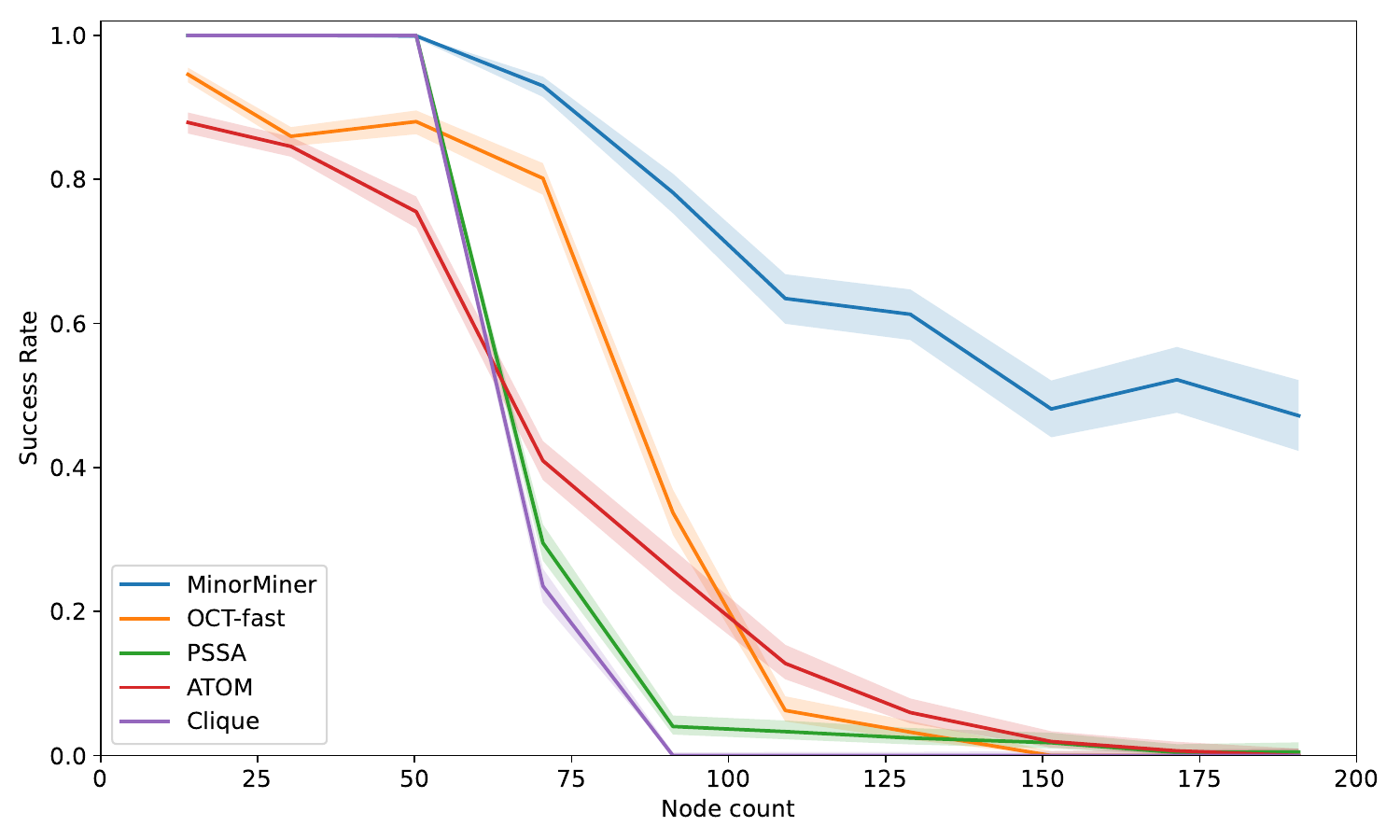}
\caption{Success rate vs.\ node count (20-node bins, all graph 
categories pooled). Shaded bands show pointwise 95\% confidence intervals (Wilson score interval
for binomial proportions, computed per 20-node bin). \textsc{Clique} fails completely above 
$n = 64$---the maximum clique $K_{64}$ embeddable in 
Chimera~$C_{16}$---producing a 100\% $\to$ 0\% transition 
in a single step. Above 80 nodes, only \textsc{MinorMiner} is maintaining meaningful success rates.}
\label{fig:success_scaling}
\end{figure}

\begin{figure}[hbtp]
\centering
\includegraphics[width=\columnwidth]{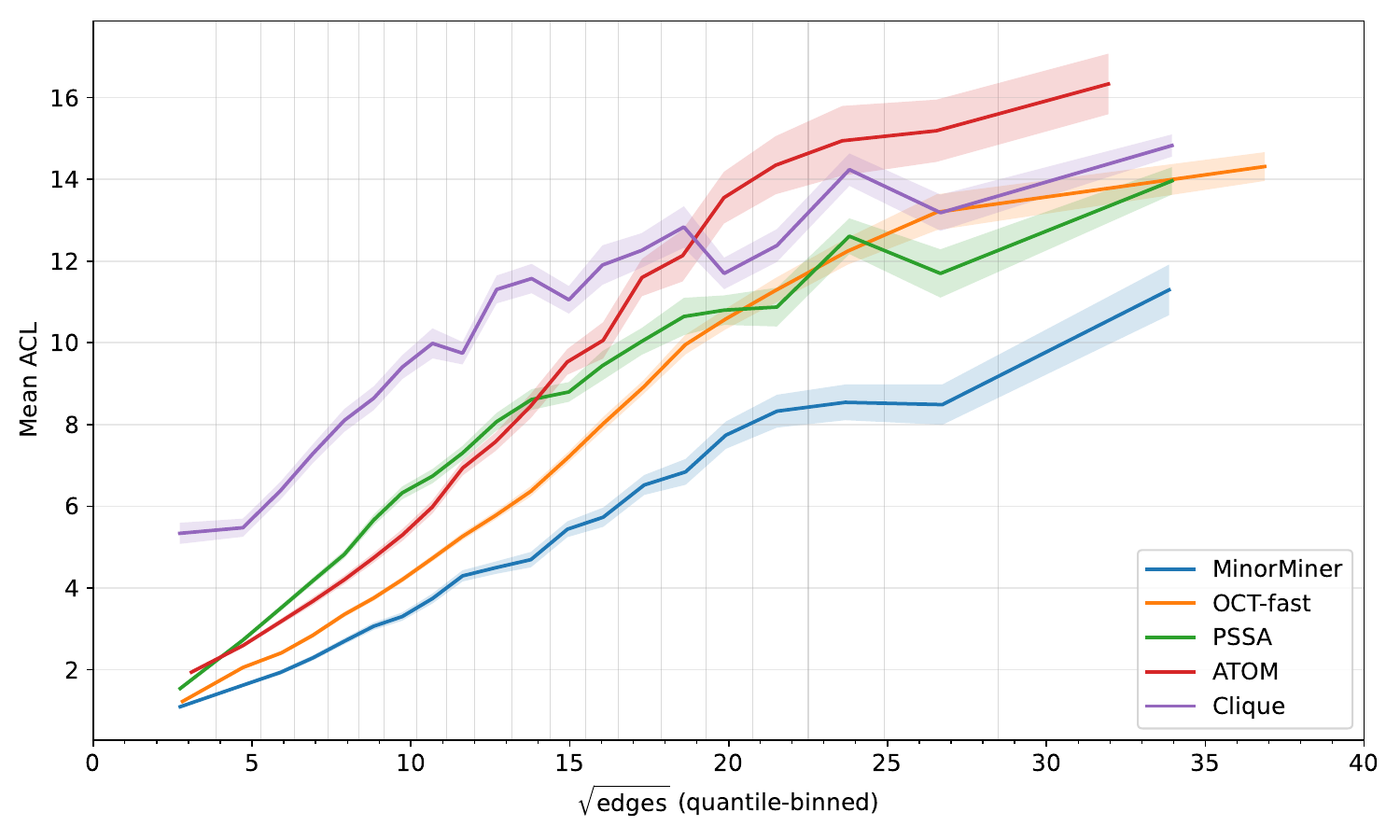}
\caption{Mean ACL vs.\ $\sqrt{|E|}$ (quantile-binned, 20 bins of
equal graph count). Shaded bands show pointwise 95\% confidence intervals for the bin mean. All algorithms exhibit approximately linear
growth; \textsc{MinorMiner} achieves the smallest slope.
\textsc{Clique}'s elevated ACL at low edge counts reflects its
$K_n$ assumption rather than edge structure.}
\label{fig:acl_scaling}
\end{figure}
\subsection{Scaling Behaviour}

\subsubsection{Success Rate Scaling}
Success rate degrades monotonically with node count (Spearman 
$\rho = -0.78$, non-linear), so we plot it against node count in Fig.~\ref{fig:success_scaling}, 
which directly determines the minimum qubit footprint of any 
embedding. Below 60 nodes all algorithms succeed on the vast 
majority of problems. \textsc{Clique} fails completely above 
$n = 64$: it allocates chains by position on the target hardware 
rather than source graph structure, bounding it by the hardware's 
maximum clique minor regardless of edge density. 
\textsc{MinorMiner} alone maintains meaningful success rates 
above 100 nodes, stabilising near 42\% at 200 nodes. Beyond 
the plotted range, \textsc{OCT-fast} produces no successful 
embeddings above $n = 128$, \textsc{ATOM} above $n = 205$, and 
\textsc{PSSA} only 9 isolated successes across the 200--700 node 
range. \textsc{MinorMiner}'s success rate decays to roughly 19\% 
at 300 nodes and 4\% at 500 nodes, with no successful embeddings 
observed for $n \geq 1{,}000$ on Chimera~$C_{16}$.

\subsubsection{ACL Scaling}
Chain length correlates most strongly with $\sqrt{|E|}$ ($r = 0.63$ 
pooled). All algorithms exhibit approximately linear ACL growth consistent with 
$\text{ACL} \approx a\sqrt{|E|} + b$, shown in Fig. \ref{fig:acl_scaling}. \textsc{MinorMiner} achieves 
the smallest slope, maintaining a 3--5 qubit advantage per chain 
that widens with edge count. At low edge counts \textsc{MinorMiner} 
and \textsc{OCT-fast} produce nearly identical chain lengths, 
diverging progressively until \textsc{OCT-fast} tracks closer to 
\textsc{PSSA} at high density which is consistent with its advantage 
being concentrated on structured rather than dense graphs. 
\textsc{ATOM} produces the longest chains among the adaptive 
algorithms despite being designed for speed, exceeding even 
\textsc{Clique} at high edge counts. \textsc{Clique} serves as 
a structural worst case: it allocates chains for a $K_n$ embedding 
regardless of actual connectivity, making it an upper bound on 
chain length for any graph of that node count.

\subsubsection{Runtime Scaling}

Embedding time splits cleanly into two regimes by algorithm 
class (log-log fits on the dominant predictor). Search-based 
algorithms are edge-driven: \textsc{MinorMiner} runtime grows 
as $|E|^{1.30}$ ($R^2 = 0.82$) and \textsc{ATOM} as 
$|E|^{1.05}$ ($R^2 = 0.82$), with node count contributing 
negligibly once edge count is controlled. Structural algorithms 
are node-driven: \textsc{OCT-fast} grows as $n^{1.3}$ 
($R^2 = 0.62$) and \textsc{PSSA} sub-linearly in nodes 
($n^{0.69}$, $R^2 = 0.58$). \textsc{Clique} runtime is 
independent of the source graph ($R^2 \approx 0$), fixed 
entirely by the target topology.

\subsection{Topology Effects}
\label{sec:topology}
We use \textsc{MinorMiner} as a single probe to isolate 
hardware-graph effects from algorithm choice; \textsc{OCT-fast}, 
\textsc{ATOM}, and \textsc{CHARME} lack Pegasus and Zephyr 
implementations.

\begin{table}[htbp]
\caption{\textsc{MinorMiner} performance by edge-count bin and 
hardware topology. Success rate is macro-averaged over all graphs; 
ACL and runtime are restricted to graphs succeeding on all three 
topologies. Bin cuts at 175 and 750 edges.}
\label{tab:topo_size}
\begin{center}
\resizebox{\columnwidth}{!}{%
\begin{tabular}{l ccc ccc ccc}
\toprule
& \multicolumn{3}{c}{\textbf{Success rate (\%)}}
& \multicolumn{3}{c}{\textbf{Mean ACL}}
& \multicolumn{3}{c}{\textbf{Median time (s)}} \\
\cmidrule(lr){2-4} \cmidrule(lr){5-7} \cmidrule(lr){8-10}
\textbf{Size}
& \textbf{Chi} & \textbf{Peg} & \textbf{Zep}
& \textbf{Chi} & \textbf{Peg} & \textbf{Zep}
& \textbf{Chi} & \textbf{Peg} & \textbf{Zep} \\
\midrule
Small  & 99.9 & 100.0 & 100.0 & 2.43 & 1.37 & 1.27 & 0.13 & 0.24 & 0.15 \\
Medium & 90.6 &  99.8 & 100.0 & 5.24 & 2.35 & 1.97 & 2.45 & 3.67 & 2.03 \\
Large  & 12.2 &  40.3 &  46.0 & 9.29 & 3.85 & 3.15 & 10.38 & 11.53 & 6.71 \\
\bottomrule
\end{tabular}}
\end{center}
\end{table}

\begin{figure*}[tp]
\centering
\includegraphics[width=\textwidth]{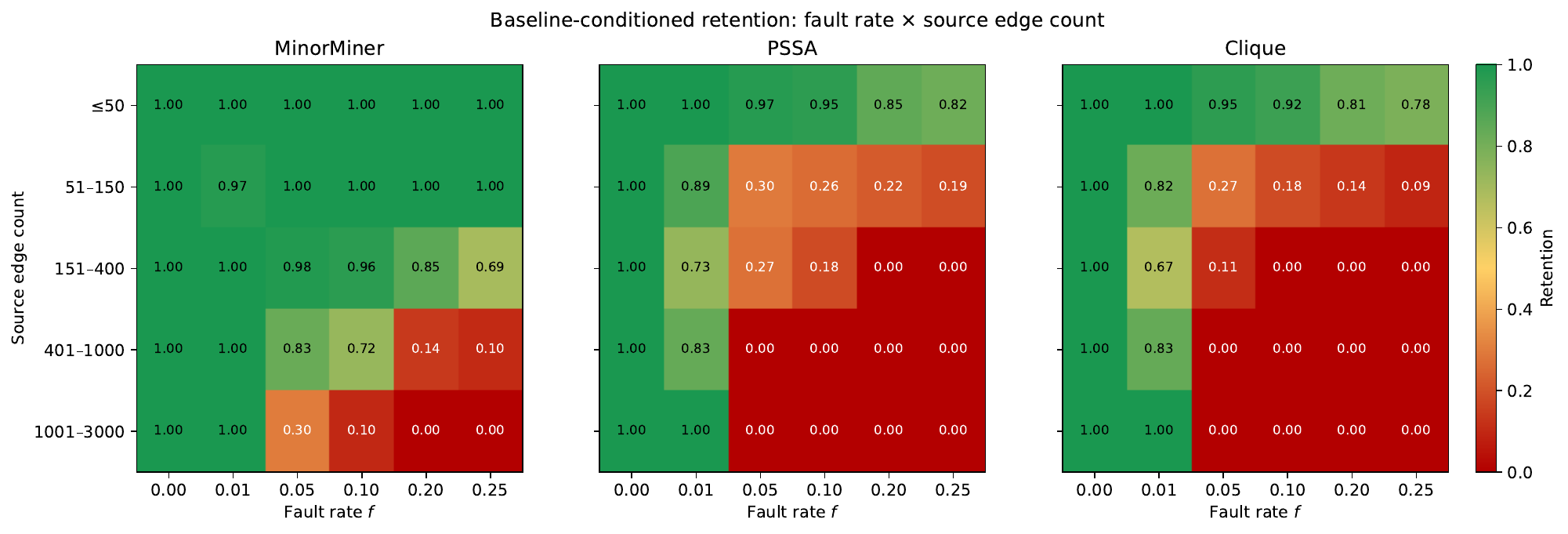}
\caption{Baseline-conditioned retention stratified by source edge count: fraction of embeddings successful at $f=0$ that remain successful at each fault rate $f$. \textsc{MinorMiner} degrades gracefully across all graph sizes; \textsc{PSSA} and \textsc{Clique} exhibit sharp cliffs between $f=0.01$ and $f=0.05$ at medium edge counts. The 1001--3000 row for \textsc{PSSA} and \textsc{Clique} is sparsely populated (few baseline successes) and its precision is correspondingly limited; the $>3000$ row has no baseline successes for any algorithm.}
\label{fig:fault_heatmap}
\end{figure*}
Table~\ref{tab:topo_size} quantifies the capacity and quality 
advantage of each topology, providing concrete thresholds for 
hardware selection. Bins are edge-count tertiles (cuts at 175 
and 750 edges) over the 15{,}307 graphs embeddable on at least 
one topology; $\log(\text{edges})$ is the stronger embedding predictor 
(pseudo-$R^2 \approx 0.75$). Zephyr achieves the shortest chains, highest success rates, 
and fastest runtimes on medium and large graphs: its higher 
connectivity (45{,}864 edges vs.\ 40{,}484 for Pegasus) means 
shorter paths between chain endpoints, and its smaller node 
count (4{,}800 vs.\ 5{,}640) reduces the search space that 
\textsc{MinorMiner} must explore per iteration. Chimera 
produces chains $2.7$--$3.0\times$ longer than Zephyr across 
the medium and large bins (5.24 vs.\ 1.97; 9.29 vs.\ 3.15), 
a direct consequence of its low node degree (${\sim}6$) forcing 
longer qubit chains to realise problem edges; this same 
connectivity bottleneck collapses its success rate to 12.2\% 
on large graphs where Pegasus and Zephyr still embed 40--46\%. 
Notably, Chimera is the fastest topology on small graphs 
(0.13\,s vs.\ 0.15\,s for Zephyr), since its compact hardware 
graph (2{,}048 nodes) reduces BFS path-finding cost; for medium 
and large graphs this advantage reverses as chain construction 
cost dominates. Pegasus is the slowest topology above the small 
bin and is ${\sim}1.7\times$ slower than Zephyr on large graphs 
(11.53\,s vs.\ 6.71\,s) despite similar success rates.

Success rate follows a clean logistic fit over $\log(\text{edges})$, 
with 50\% thresholds at 747 edges on Chimera, 2{,}273 on Pegasus, 
and 2{,}557 on Zephyr which has roughly $3.0\times$ and $3.4\times$ 
more edge capacity than Chimera, with Zephyr extending Pegasus's 
capacity by a further 12\%.

\subsection{Fault Tolerance}
\label{sec:fault}

Real quantum hardware operates with persistent qubit and coupler 
faults that every deployed embedding must tolerate~\cite{lobe2021broken,
dwave2020advantage,pelofske2025}. Early work by Klymko et al.~\cite{klymko2014} established that
fault-resilient embedding algorithms significantly outperform naive
approaches as fabric fault rates increase, motivating the fault
tolerance analysis we conduct here across realistic D-Wave fault rates. We simulated faults on Chimera~$C_{16}$ at rates 
$f \in \{0.00, 0.01, 0.05, 0.10, 0.20, 0.25\}$, where $f$ is the 
fraction of qubits removed: at each $f$, exactly $\lfloor N \cdot f 
\rfloor$ nodes are drawn without replacement uniformly at random 
from the $N$-node hardware graph and removed prior to embedding, 
along with all incident edges. Uniform random removal does not 
reproduce the spatial correlations of real hardware fault maps, so 
results indicate general algorithmic robustness rather than 
deployed performance; Ember also supports explicit fault-pattern 
specification for benchmarking against real QPU calibration 
snapshots. We measured both absolute success rate and 
\emph{retention}: the fraction of baseline-successful embeddings 
($f = 0$) that remain successful at each $f$. \textsc{OCT-fast}, 
\textsc{ATOM}, and \textsc{CHARME} are excluded, since fault 
simulation requires modifying the target topology and their 
implementations are Chimera-bound.  

\textsc{MinorMiner} degrades gracefully with rising fault rates, 
losing 28\% of its baseline successes at $f=0.25$. \textsc{PSSA} 
and \textsc{Clique} each lose roughly twice that fraction (49\% 
relative) over the same range. The gap is concentrated in a sharp 
cliff between $f=0.01$ and $f=0.05$: \textsc{PSSA} and \textsc{Clique} 
drop $\sim$12\,pp of absolute success in that single step, while 
\textsc{MinorMiner} drops 4\,pp. Fig.~\ref{fig:fault_heatmap} 
localises the cliff to the medium-edge regime (151--400 edges), 
where \textsc{MinorMiner} retains $0.98$ of its baseline embeddings 
and \textsc{PSSA} and \textsc{Clique} retain $0.27$ and $0.11$ 
respectively. The largest-edge row is sparsely populated at baseline 
and should be read as a lower bound.

Chain quality for successful embeddings is flat in \textsc{MinorMiner} 
(mean ACL 3.30$\rightarrow$3.17 across the fault range). 
\textsc{Clique}'s mean ACL falls sharply (7.74$\rightarrow$3.54), 
but this is a survivorship effect: only smaller source graphs 
continue to embed at high fault rates which have shorter chains 
regardless of algorithm. Runtime grows for search-based algorithms 
(\textsc{MinorMiner} $+65\%$, \textsc{PSSA} $+41\%$ across the range) 
as constraint satisfaction becomes harder with fewer available qubits. 
The underlying mechanism is visible: \textsc{MinorMiner}'s incremental 
path-finding routes around isolated faults, whereas \textsc{Clique}'s 
fixed template and \textsc{PSSA}'s reliance on globa
target-graph properties both degrade discontinuously. 
\textsc{MinorMiner}'s advantage on ideal topologies therefore widens 
further on real hardware, where fault rates are non-zero.

\subsection{CHARME: Learned Ordering at the Training Distribution}
\label{sec:charme}

We evaluated \textsc{CHARME} on all 8,891 Chimera, embeddable 
graphs with $n \leq 120$ in the main Ember library, the subset CHARME's architecture can run on via zero-padding 
(\S\ref{sec:algorithms_evaluated}). Because few graphs at exactly $n = 120$ in Ember 
are embeddable on Chimera($16,16,4$) by any algorithm at this 
size, we supplemented with a 45-graph diagnostic of $n = 120$ 
graphs spanning 14 categories with 5 trials per graph 
(Table~\ref{tab:charme_summary}), giving \textsc{CHARME} a 
representative test set at its native size.

\begin{table}[hbtp]
\centering
\caption{Success rate by algorithm partitioned by node count. 
The first three columns cover $n < 120$ from the main Ember library; the last column is the 45-graph custom suite at $n = 120$ 
(\textsc{CHARME}'s native size, no padding). Note 
\textsc{CHARME}'s discontinuity between the $75 \leq n < 120$ 
column (24\%) and the $n = 120$ column (51\%).}
\label{tab:charme_summary}
\begin{tabular}{lcccc}
\toprule
 & \multicolumn{3}{c}{padded (\textsc{CHARME} pads to 120)} & native \\
\cmidrule(lr){2-4}\cmidrule(lr){5-5}
Algorithm & $n < 25$ & $25 \le n < 75$ & $75 \le n < 120$ & $n = 120$ \\
\midrule
MinorMiner       & 100\% & 99\% & 75\% & 100\% \\
\textbf{CHARME}  & \textbf{88\%} & \textbf{69\%} & \textbf{24\%} & \textbf{51\%} \\
ATOM             & 88\%  & 73\% & 22\% & 29\% \\
OCT-fast         & 94\%  & 86\% & 29\% & 20\% \\
PSSA             & 100\% & 87\% &  4\% & 19\% \\
Clique           & 100\% & 86\% &  0\% &  0\% \\
\bottomrule
\end{tabular}
\end{table}

The jumps from the $75 \leq n < 120$ column to the $n = 120$ 
column reflect a change in test-set composition (the 45-graph 
diagnostic is sparser than the average graph in the $75$--$119$ 
bin), and several algorithms show modest improvements between 
the two columns. \textsc{CHARME}'s jump, however, is larger than 
any other algorithm's and is most cleanly seen relative to 
\textsc{ATOM}, its mechanical backbone: the two algorithms track 
each other across all three padded columns (within 4 pp at every 
$n$ bin), but at $n = 120$ \textsc{CHARME} jumps to 22 pp ahead, 
moving from fifth place to second behind only \textsc{MinorMiner}. 
The learned ordering's contribution over the heuristic backbone 
emerges sharply at the training distribution and is negligible 
elsewhere.

\textsc{CHARME}'s success-rate gain at $n = 120$ comes at a substantial 
embedding-quality penalty consistent across both evaluation suites. 
On the $5{,}518$ graphs where both \textsc{CHARME} and 
\textsc{ATOM} succeed, \textsc{CHARME}'s 
chains are longer in $81.7\%$ of cases---mean ACL $+34\%$ and maximum chain length $+46\%$. Against \textsc{MinorMiner} the penalty is uniform: ACL is $1.96\times$ and MCL $2.94\times$ higher. Runtime scales linearly with $n$, putting \textsc{CHARME} 
at ${\sim}20\times$ \textsc{ATOM}'s wall time across both regimes; 
against \textsc{MinorMiner}, \textsc{CHARME} is $4.8\times$ slower 
at $n = 120$ and $1.25\times$ slower for  $n < 120$. The learned 
policy trades chain quality for narrowly scoped feasibility gains: it extends 
coverage beyond \textsc{ATOM}'s by accepting longer chains and more 
qubits, but is still uniformly worse than \textsc{MinorMiner} on every 
embedding-quality axis.

\section{Discussion}

\subsection{Practical Recommendations}

The results support concrete guidance for practitioners selecting
embedding algorithms. Three of the five evaluated algorithms
(\textsc{OCT-fast}, \textsc{ATOM}, \textsc{CHARME}) currently
target Chimera only, leaving \textsc{MinorMiner}, \textsc{Clique},
and \textsc{PSSA} as the options on Pegasus and Zephyr; of these,
\textsc{MinorMiner} dominates on all key metrics (\S\ref{sec:topology}).
For Chimera, where a full five-algorithm comparison was possible,
\textsc{MinorMiner} is the strongest default choice across the
broadest range of inputs and remains the only algorithm with
substantial success rates above 100 nodes. However, aggregate
performance is an incomplete guide. The benchmark exposes several
findings worth flagging:

\textbf{Match algorithm to graph structure when known.} 
\textsc{OCT-fast} produces chains 7.8--15.8\% shorter than 
\textsc{MinorMiner} on algebraically regular graphs (Tur\'an, 
Kneser, hypercube, bipartite) and on spin-glass instances; 
\textsc{PSSA} leads \textsc{MinorMiner} on complete graphs; 
\textsc{Clique} is near-optimal for complete-graph inputs within 
Chimera's $K_{64}$ capacity. \textsc{MinorMiner} leads on every 
other category, with strongest dominance on physics lattices and 
topological primitives.

\textbf{\textsc{OCT-fast} is the practical first try for 
runtime-sensitive workloads.} The conventional framing of 
\textsc{OCT} as a slower, higher-quality alternative is inverted: 
when \textsc{OCT-fast} succeeds, it produces chains within 3\% of 
\textsc{MinorMiner}'s at roughly one-eighth the runtime. One may consider trying it 
first and falling back to \textsc{MinorMiner} on failure.

\textbf{\textsc{ATOM}'s speed advantage does not justify its 
quality cost.} \textsc{ATOM} runs at ${\sim}28\times$ 
\textsc{MinorMiner}'s speed and roughly $3\times$ 
\textsc{OCT-fast}'s, but pays for it with both lower success 
rate and longer chains than \textsc{OCT-fast} on this benchmark. 
For practitioners who would otherwise choose \textsc{ATOM} for 
its speed, \textsc{OCT-fast} is the better tradeoff: it is still 
much faster than \textsc{MinorMiner} while significantly outperforming 
\textsc{ATOM} on quality.

\textbf{Hardware leverage exceeds algorithm leverage on large 
problems.} \textsc{MinorMiner} embeds 12.2\% of large graphs on 
Chimera versus 46.0\% on Zephyr showing a $3.8\times$ improvement from 
hardware alone, larger than any algorithm-side gain within Chimera. 
Chain quality follows the same pattern: mean ACL on 
large graphs drops from 9.29 (Chimera) to 3.85 (Pegasus) to 3.15 
(Zephyr), with most of the gain captured in the Chimera-to-Pegasus 
step.

\textbf{Algorithm robustness diverges at low fault rates.} 
\textsc{PSSA} and \textsc{Clique} lose roughly three times more 
absolute success than \textsc{MinorMiner} ($\sim$12\,pp vs 
$4$\,pp) at fault rates inside typical D-Wave operating ranges. 
\textsc{MinorMiner}'s lead widens on real hardware rather than 
narrows.

\subsection{Limitations}
\paragraph{Embedding quality as a proxy for annealing performance}
The benchmark evaluates embeddings independently of QPU solution 
quality. Chain length is an established proxy for annealing 
fidelity~\cite{venturelli2015,gomez2025}, but the mapping is 
problem-dependent. End-to-end evaluation is left for future work.

\paragraph{Algorithm--topology coverage asymmetry}
Three of the six evaluated algorithms (\textsc{OCT}, \textsc{ATOM}, 
\textsc{CHARME}) have published implementations targeting Chimera 
only, restricting the main algorithm comparison (\S\ref{sec:overall}, \S\ref{sec:by_family}) to 
Chimera and forcing topology analysis (\S\ref{sec:topology}) to use 
\textsc{MinorMiner} as a single probe. Our \textsc{PSSA} 
reimplementation, which generalises the original Hitachi-CMO 
design to D-Wave's three topologies, demonstrates that this 
generalisation is feasible but non-trivial. Until \textsc{OCT}, 
\textsc{ATOM}, and \textsc{CHARME} are similarly extended, 
cross-algorithm comparisons on Pegasus, Zephyr and varying fault rates remain 
out of reach.

\paragraph{Fault simulation methodology}
Uniform random qubit removal does not reproduce the spatial 
correlations of real hardware fault maps. The fault tolerance 
results (\S\ref{sec:fault}) indicate general algorithmic robustness rather 
than predicting deployed performance on a specific QPU. Ember 
supports explicit fault-pattern specification for benchmarking 
against real calibration snapshots.

\paragraph{CHARME reimplementation}
Three constraints bound \textsc{CHARME}'s results. The fixed 
action dimension limits per-checkpoint coverage to graphs at or 
below the trained size (via padding). The original \textsc{ATOM} binary 
used for chain construction during training fails on a fraction 
of episodes due to memory and concurrency issues, reducing 
effective training throughput. The order exploration strategy 
(Algorithm 2 in~\cite{ngo2024charme}) could not be fully 
implemented due to \textsc{ATOM}'s parallel file-write conflicts; 
identity ordering was used instead, which the original paper 
shows slows convergence (Fig.~7 in~\cite{ngo2024charme}). 
Pretrained weights were not released, so our reimplementation 
is not validated against the published evaluation; absolute 
numbers should be read as indicative, while qualitative patterns 
follow from the algorithm's construction.

\paragraph{PSSA reimplementation}
No publicly available implementation generalised PSSA to all
D-Wave topologies prior to this work; our reimplementation
adapts the original Hitachi-CMO
design~\cite{sugie2018,sugie2021} with three substantive
changes: busclique-derived initialisation replacing the 
King's-graph guiding path, tree-shaped chain mutations replacing 
simple-path mutations, and per-topology annealing schedule 
rescaling (\S\ref{sec:algorithms_evaluated}). These adaptations were necessary to produce 
a functioning implementation on Chimera, Pegasus, and Zephyr, 
but they mean the results labelled \textsc{PSSA} throughout this 
paper reflect our reimplementation rather than the original 
algorithm. Without access to the original codebase for 
validation, we cannot quantify the performance gap, if any, 
between the two. Qualitative behaviour---simulated-annealing 
search over chain configurations with terminal-search 
post-processing---is preserved from the original paper.


\section{Conclusions and Future Work}

We have presented Ember, an open-source benchmarking framework 
for quantum annealing minor embedding algorithms, installable with a single \texttt{pip install} and prepackaged with a variety of state-of-the-art minor embedding algorithms.
Ember addresses three gaps in the existing literature: the absence of a 
standardized platform enabling cross-algorithm comparison, the 
exclusive reliance on random graph types that do not reflect 
the structural diversity of practical quantum annealing 
workloads, and insufficient evaluation under deployment-relevant 
conditions including faulty hardware and varying topology.

Across 17{,}248 Chimera-embeddable graphs spanning 35 categories, 
our finding that relative algorithm performance varies 
systematically across graph structure helps gives a fuller picture 
of the algorithm landscape. \textsc{ATOM} was reported in its 
original evaluation as a faster alternative to \textsc{MinorMiner} 
on Barab\'asi--Albert and $d$-regular graphs, but our broader 
evaluation shows it is dominated on several quality metrics by 
\textsc{OCT-fast}, which itself was not previously compared 
against \textsc{ATOM} on common ground. \textsc{MinorMiner} was 
not previously evaluated on physics-lattice or 
topological-primitive graphs central to real quantum simulation 
workloads, where Ember shows it dominates. 

Using \textsc{MinorMiner} as a probe, Ember further showed that hardware topology gains---Zephyr extending Chimera's effective embedding capacity by roughly $3\times$-can exceed algorithm-side improvements on large problems, and that
\textsc{MinorMiner}, \textsc{PSSA}, and \textsc{Clique} diverge sharply in robustness to qubit faults at deployment-relevant rates.


Obvious next steps include generalizing \textsc{OCT}, \textsc{ATOM}, and \textsc{CHARME} to Pegasus and Zephyr; Ember's algorithm contract provides a stable interface for community contributions, and our \textsc{PSSA} reimplementation illustrates the engineering pattern. 
Topology-generalised \textsc{OCT-fast} is particularly high-value 
given its near-\textsc{MinorMiner} chain quality on Chimera 
combined with Zephyr's higher connectivity. Within \textsc{OCT}, 
the order-of-magnitude variant spread observed in \S\ref{sec:variant_selection} points 
to unexplored structural design space; targeted refinement of 
decomposition and chain-construction procedures could plausibly 
close the remaining gap to \textsc{MinorMiner} or potentially outperform it. For learned-ordering approaches, graph-conditioned policies handling 
variable input sizes natively address the binding architectural 
constraint identified in \textsc{CHARME}'s evaluation.

Beyond topology generalisation, Ember's algorithm contract
enables evaluation of additional embedding approaches not
covered here, including integer-programming methods~\cite{bernal2020}, layout-aware embedding~\cite{pinilla2019}, and specialised structural
embeddings~\cite{sinno2025}, broadening the comparison as the algorithm ecosystem grows.
One may also consider a natural extension of \textsc{CHARME}: replacing \textsc{ATOM}
as the chain-construction backbone with different algorithms such as \textsc{MinorMiner} or
\textsc{OCT-fast}, which our results show dominate \textsc{ATOM}
on quality; this could preserve the learned ordering's
feasibility advantage while inheriting superior chain quality.

Additional benchmark extensions are also possible. First, enriching the 
graph library with quantitative structural metadata (treewidth, 
clustering coefficient, degree distribution moments, spectral 
properties) to test whether algorithm selection can be driven 
by measured graph properties rather than category labels, 
replacing ``match algorithm to graph type'' with ``match algorithm 
to graph features.'' Second, end-to-end evaluation connecting 
embedding quality to annealing solution quality on real D-Wave 
hardware, closing the chain-length-as-proxy gap. Third, 
fault-pattern benchmarking against real QPU calibration snapshots 
rather than uniform random removal.

Lastly, one could consider testing embedding algorithms in the context of networked quantum devices, in combination with graph/problem decomposition algorithms \cite{pelofske2021decomposition,krpan2025quantum}.





\end{document}